\documentclass[pra,twocolumn]{revtex4}%
\usepackage{amsfonts}
\usepackage{amsmath}
\usepackage{amssymb}
\usepackage{graphicx}%
\providecommand{\U}[1]{\protect\rule{.1in}{.1in}}

\begin{document}
\title{Frequency-up conversion and quantum swap gate in an optical cavity with atomic cloud}
\author{Gong-Wei Lin$^{1}$}
\author{Xu-Bo Zou$^{2}$}
\email{xbz@ustc.edu.cn}
\author{Ming-Yong Ye$^{1}$}
\author{Xiu-Min Lin$^{1}$}
\email{xmlin@fjnu.edu.cn}
\author{Guang-Can Guo$^{2}$}
\affiliation{$^{1}$School of Physics and Optoelectronics Technology, Fujian Normal
University, Fuzhou 350007, People's Republic of China}
\affiliation{$^{2}$Key Laboratory of Quantum Information, Department of Physics, University
of Science and Technology of China, Hefei 230026}

\begin{abstract}
A scheme is presented for realizing frequency-up conversion and a two-qubit
quantum swap gate for intracavity fields. In the scheme, a V-type atomic
ensemble prepared in their ground states collectively mediates the interaction
between the two cavity modes. Under certain conditions, the cavity-field
degree of freedom is decoupled from the atomic degrees of freedom, and the
effective coupling strength between the two cavity modes scales up with
$\sqrt{N}$ ($N$ is the number of atoms). The numerical simulation shows that
the quantum swap gate still has a high fidelity under the influence of the
atomic spontaneous emission and the decay of the cavity modes.

\end{abstract}

\maketitle

\section{Introduction}

Cavity quantum electrodynamics (QED) devices hold great promise as basic tools
for quantum networks \cite{Cirac} since they provide an interface between
computation and communication, i.e., between atoms and photons. In this
context, it is a very important task to coherently manipulate and control
quantum state of not only the atom but also the intracavity fields. Throughout
the last decade, in cavity QED including the microwave and optical regimes
many efforts have been devoted to generation of entangled states, nonclassical
states, and the implementation of quantum logical gates for cavity fields.
Some relevant examples are Schr\"{o}dinger cat state \cite{Brune}, Fock state
of a single-mode cavity field \cite{Brattke}, the maximally entangled state of
two cavity modes \cite{Rauschen}, pair coherent states and $SU(2)$ coherent
states \cite{Zheng}. In fact, three different schemes were proposed to realize
a quantum phase gate of two intracavity modes in which a single atom with
$\Xi$-type \cite{Solano}, $V$- type \cite{Garcia}, or $\Lambda$-type
\cite{Shu} configuration mediated the interaction between them. Recently,
Serra et al. \cite{Serra} and Zou et al. \cite{Zou} showed the methods for
realizing frequency up- and down-conversions in a two-mode microwave cavity
based on dispersive interaction and resonant interaction, respectively.

Most of the schemes mentioned above are based on the interaction of single
atoms and the cavity fields. In the past few years, many interesting schemes
have been proposed that use atomic ensembles with a large number of identical
atoms placed in an optical cavity as the basic system for quantum state
engineering and quantum information processing \cite{Sondb,Lukin,Dantan,Duan}.
In particular, several experiments have been reported about continuous
variable entanglement using cold atoms \cite{Josse} and the photon statistics
of the light emitted from an atomic ensemble into a single cavity mode
\cite{Hennrich}. Theoretically, some schemes have also been proposed for
realizing frequency down-conversions--- two-mode field squeezing using atomic
ensemble as medium \cite{Guzman} and carrying out quantum phase-gate operation
for two single photons using an $N$-type atomic ensemble trapped in an optical
cavity \cite{Xiao} or an M-type atomic ensemble trapped in a gas cell
\cite{Ottaviani}. The above schemes based on atomic ensembles have some
special advantages compared with those based on the control of single atoms:
first, laser manipulation of atomic ensembles without separately addressing
the individual atoms is normally easier than the coherent control of single
particles; second, and more important, atomic ensembles could have some kinds
of collectively enhanced coupling to certain optical mode due to the many-atom
interference effects.

In this paper, we present a method that realizes frequency up-conversions and
a quantum swap gate for two cavity modes with a V-type atomic ensemble in an
optical cavity. In the scheme, the cavity-field degree of freedom is decoupled
from the atomic degrees of freedom and the effective coupling strength scales
up with $\sqrt{N}$ (here $N$ is the number of atoms). The scheme is robust
against decoherence processes, such as spontaneous emission, and does not
require a strong coupling regime or precise atomic localization. The numerical
simulation shows that the swap gate has high fidelity and small error rate
under the influence of the atomic spontaneous emission and the decay of the
cavity modes.

\section{The fundamental model and realization of frequency-up conversion}

Our model consists of an ensemble of N identical V-type atoms inside a
two-mode optical cavity. The concrete atomic level structure and relevant
transitions are shown in Fig. 1. \begin{figure}[ptb]
\includegraphics[width=1.8in]{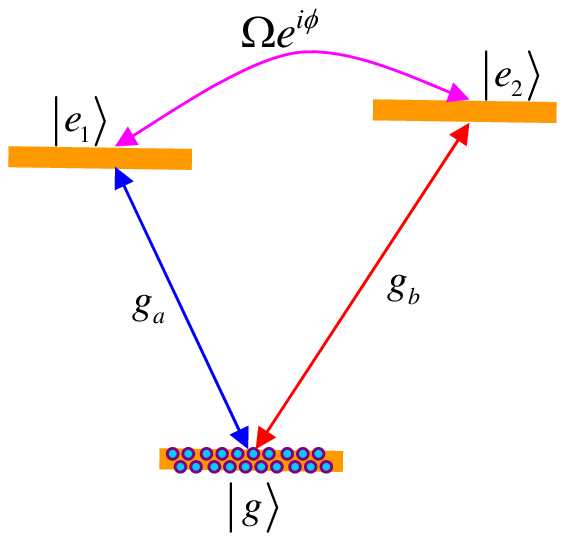}\newline\caption{(Color online)
Configuration of the atomic level structure and relevant transitions. The
states $\left\vert e_{1}\right\rangle $ and $\left\vert e_{2}\right\rangle $
correspond to two excited states\ while $\left\vert g\right\rangle $ is a
stable ground state. The transitions $\left\vert g\right\rangle
\leftrightarrow$ $\left\vert e_{1}\right\rangle $ and $\left\vert
g\right\rangle \leftrightarrow$ $\left\vert e_{2}\right\rangle $ of each atom
couple to the cavity modes $a$ and $b$ with coupling strengths $g_{a}$ and
$g_{b}$, respectively. Simultaneously the external laser field with Rabi
frequency $\Omega e^{i\phi}$ drives the Raman transition $\left\vert
e_{1}\right\rangle \leftrightarrow\left\vert e_{2}\right\rangle $.}%
\label{a}%
\end{figure}Each atom has a stable ground state $\left\vert g\right\rangle $
and two excited states $\left\vert e_{1}\right\rangle $ and $\left\vert
e_{2}\right\rangle $. The transition $\left\vert g_{j}\right\rangle
\leftrightarrow$ $\left\vert e_{j1}\right\rangle $ ($\left\vert g_{j}%
\right\rangle \leftrightarrow$ $\left\vert e_{j2}\right\rangle $) of the jth
atom couples to the cavity mode $a$ ($b$) with coupling strength $g_{ja}$
($g_{jb}$), while laser field with Rabi frequency $\Omega_{jj}e^{i\phi_{j}}$
drives the transition $\left\vert e_{j1}\right\rangle \leftrightarrow$
$\left\vert e_{j2}\right\rangle $ ($\Omega_{jj}e^{i\phi_{j}}$ can be realized
by using a two-photon Raman transition via level $\left\vert g_{j}%
\right\rangle $, or a fourth level \cite{Jeremy}). The Hamiltonian for the
whole system in the interaction picture can be written as%

\begin{equation}
H_{I}=H_{cav}+H_{cla},
\end{equation}
with $H_{cav}=\sum\limits_{j=1}^{N}(g_{ja}a\left\vert e_{j1}\right\rangle
\left\langle g_{j}\right\vert +g_{jb}b\left\vert e_{j2}\right\rangle
\left\langle g_{j}\right\vert )+H.c.$ and $H_{cla}=\sum\limits_{j=1}^{N}%
\Omega_{j}e^{i\phi_{j}}\left\vert e_{j2}\right\rangle \left\langle
e_{j1}\right\vert +H.c.$.

After defining the new atomic basis $\left\vert +_{j}\right\rangle
=($\ $\left\vert e_{j1}\right\rangle +e^{i\phi_{j}}\left\vert e_{j2}%
\right\rangle )/\sqrt{2}$, $\left\vert -_{j}\right\rangle =($\ $\left\vert
e_{j1}\right\rangle -e^{i\phi_{j}}\left\vert e_{j2}\right\rangle )/\sqrt{2}$
\cite{Zhen}, one can rewrite $H_{cav}$ and $H_{cla}$ as $H_{cav}%
=\sum\limits_{j=1}^{N}(\frac{g_{ja}}{\sqrt{2}}a(\ \left\vert +_{j}%
\right\rangle +\left\vert -_{j}\right\rangle )\left\langle g_{j}\right\vert
+\frac{g_{jb}e^{-i\phi_{j}}}{\sqrt{2}}b(\ \left\vert +_{j}\right\rangle
-\left\vert -_{j}\right\rangle )\left\langle g_{j}\right\vert )+H.c.$ and
$H_{cla}=\sum\limits_{j=1}^{N}\Omega_{j}(\left\vert +\right\rangle
_{j}\left\langle +\right\vert -\left\vert -\right\rangle _{j}\left\langle
-\right\vert )$. The time evolution of this system is decided by
Schr\"{o}dinger's equation%

\begin{equation}
i[d\left\vert \Psi(t)\right\rangle /dt]=H_{I}\left\vert \Psi(t)\right\rangle .
\end{equation}
After performing the unitary transformation%

\begin{equation}
\left\vert \Psi(t)\right\rangle =e^{-iH_{cla}t}\left\vert \Psi^{^{\prime}%
}(t)\right\rangle ,
\end{equation}
one can obtain%

\begin{equation}
i[d\left\vert \Psi^{^{\prime}}(t)\right\rangle /dt]=H_{II}\left\vert
\Psi^{^{\prime}}(t)\right\rangle ,
\end{equation}
where%

\begin{align}
H_{II}  &  =\sum\limits_{j=1}^{N}(\frac{g_{ja}}{\sqrt{2}}a(\ e^{i\Omega_{j}%
t}\left\vert +_{j}\right\rangle +e^{-i\Omega_{j}t}\left\vert -_{j}%
\right\rangle )\left\langle g_{j}\right\vert \nonumber\\
&  +\frac{g_{jb}e^{-i\phi_{j}}}{\sqrt{2}}b(\ e^{i\Omega_{j}t}\left\vert
+_{j}\right\rangle -e^{-i\Omega_{j}t}\left\vert -_{j}\right\rangle
)\left\langle g_{j}\right\vert )+H.c..
\end{align}

Assuming that all atoms are initially in the ground state $\prod_{j=1}%
^{N}\left\vert g_{j}\right\rangle $, and%

\begin{equation}
\Omega_{j}\gg\sqrt{N}\left\vert g_{ja}\right\vert ,\sqrt{N}\left\vert
g_{jb}\right\vert ,
\end{equation}
we neglect the effect of rapidly oscillating terms. Using the time-averaging
method of Ref. \cite{Zubair,James}, we can further reduce $H_{II}$ to an
effective interaction Hamiltonian%

\begin{equation}
H_{eff}=-(\xi a^{\dagger}b+\xi^{\ast}b^{\dagger}a)\prod_{j=1}^{N}\left\vert
g_{j}\right\rangle \left\langle g_{j}\right\vert ,
\end{equation}
where $\xi=\sum\limits_{j=1}^{N}\frac{g_{ja}^{\ast}g_{jb}e^{-i\phi_{j}}%
}{\Omega_{j}}$ is the effective coupling constant. Eq. (7) is the expected
frequency up-conversion process, which belongs to a group of nonlinear
phenomena of great practical importance and also has very interesting
theoretical aspects \cite{1}. Compared with the previous theoretical protocols
for realizing frequency up-conversion with an atom in a microwave cavity
\cite{Serra,Zou}, the effective coupling constant $\xi$ in our scheme scales
with $\sqrt{N}$ though the Rabi frequency of laser field $\Omega_{j}$ should
satisfy the conditions in Eq. (6). In principle we can obtain a desired
coupling strength as long as enough atoms simultaneously interact with the
cavity modes and laser field.

\section{Quantum swap gate}

Next we will show in detail the implementation of a quantum swap gate based on
Eq. (7). Suppose that the zero- and one-photon Fock states of two different
polarization (or different frequency) modes of the radiation field inside the
cavity represent two logical states of a qubit. When the atoms are initially
prepared in their ground states $\prod_{j=1}^{N}\left\vert g_{j}\right\rangle
$ through optical pumping \cite{Duan} and are driven by the two cavity modes
and the appropriate laser field (Fig. 1), the time evolution of four logical
states for two qubits, under the effective Hamiltonian in Eq. (7), are given by%

\begin{align}
\left\vert 0\right\rangle \left\vert 0\right\rangle  &  \rightarrow\left\vert
0\right\rangle \left\vert 0\right\rangle ,\nonumber\\
\left\vert 0\right\rangle \left\vert 1\right\rangle  &  \rightarrow\cos(\xi
t)\left\vert 0\right\rangle \left\vert 1\right\rangle +i\sin(\xi t)\left\vert
1\right\rangle \left\vert 0\right\rangle ,\nonumber\\
\left\vert 1\right\rangle \left\vert 0\right\rangle  &  \rightarrow\cos(\xi
t)\left\vert 1\right\rangle \left\vert 0\right\rangle +i\sin(\xi t)\left\vert
0\right\rangle \left\vert 1\right\rangle ,\nonumber\\
\left\vert 1\right\rangle \left\vert 1\right\rangle  &  \rightarrow\cos(2\xi
t)\left\vert 1\right\rangle \left\vert 1\right\rangle +i\sin(2\xi
t)(\left\vert 2\right\rangle \left\vert 0\right\rangle +\left\vert
0\right\rangle \left\vert 2\right\rangle )/\sqrt{2},
\end{align}
where we have assumed $\xi$ is real without loss of generality.

\bigskip After the effective interaction time $t=\frac{\pi}{2\xi}$, one has%

\begin{align}
\left\vert 0\right\rangle \left\vert 0\right\rangle  &  \rightarrow\left\vert
0\right\rangle \left\vert 0\right\rangle ,\text{ \ }\left\vert 0\right\rangle
\left\vert 1\right\rangle \rightarrow i\left\vert 1\right\rangle \left\vert
0\right\rangle ,\nonumber\\
\left\vert 1\right\rangle \left\vert 0\right\rangle  &  \rightarrow
i\left\vert 0\right\rangle \left\vert 1\right\rangle ,\text{ }\left\vert
1\right\rangle \left\vert 1\right\rangle \rightarrow-\left\vert 1\right\rangle
\left\vert 1\right\rangle ,
\end{align}
which represent a swap gate operation, apart from phase factors that can be
eliminated by an appropriate setting of the phase of subsequent logic
operations \cite{Barenc}. We note that the swap gate is not a required
composition of elementary gates from a universal set for quantum computation,
which can be decomposed into three quantum phase gates and six Hadamard gates
\cite{Vatan,Sangouar}. Our aim for the direct construction of a specific
gate-- swap gate is to reduce the number of required physical logical gates in
practical quantum information processes. We also note that Sangouard et al.
\cite{Vatan} and Yavuz \cite{Yavuz} have proposed the scheme for direct swap
gate for the atoms by adiabatic passage and for the photons by using
electromagnetically induced transparency, respectively.

In the following, we analyze the fidelity of the swap gate and the photon loss
during the gate operation. We suppose that no photon is detected either by the
atomic spontaneous emission or by the leakage of a photon through the cavity
mirrors. The evolution of the system is governed by the non-Hermitian
Hamiltonian \cite{Plenio}%

\begin{equation}
H_{I}^{^{\prime}}=H_{I}-\frac{i}{2}\sum\limits_{j=1}^{N}\sum\limits_{m=1}%
^{2}\gamma_{m}\left\vert e_{jm}\right\rangle \left\langle e_{jm}\right\vert
-\frac{i}{2}(\kappa_{a}a^{\dagger}a+\kappa_{b}b^{\dagger}b),
\end{equation}
where $\kappa_{a}$ and $\kappa_{b}$ denote the decay rate of the cavity mode
fields a and b, $\gamma_{m}$ is spontaneous emission rate of the excited state
$\left\vert e_{m}\right\rangle $. For the sake of convenience, we assume
$\kappa_{a}=\kappa_{b}=\kappa$, $\gamma_{1}=\gamma_{2}=\gamma_{s}$,
$\kappa=\gamma_{s}$, $g_{ja}=g_{jb}=g$, and $\Omega_{j}=\Omega$ in the
following numerical simulations \cite{11}, so it can be seen $\xi
=\frac{N\left\vert g\right\vert ^{2}}{\Omega}$. If the initial state of the
atom-cavity system is prepared in $\left\vert \Psi(0)\right\rangle
=\prod_{j=1}^{N}\left\vert g_{j}\right\rangle \otimes(\left\vert
0\right\rangle \left\vert 0\right\rangle +\left\vert 0\right\rangle \left\vert
1\right\rangle +\left\vert 1\right\rangle \left\vert 0\right\rangle
+\left\vert 1\right\rangle \left\vert 1\right\rangle )/2$, after the effective
interaction time $t=\frac{\pi}{2\xi}$ under the non-Hermitian Hamiltonian in
Eq. (10), the conditional state can be described by $\left\vert \Psi
(t_{I})\right\rangle =\prod_{j=1}^{N}\left\vert g_{j}\right\rangle
\otimes(\alpha_{00}\left\vert 0\right\rangle \left\vert 0\right\rangle
+\alpha_{01}\left\vert 0\right\rangle \left\vert 1\right\rangle +\alpha
_{10}\left\vert 1\right\rangle \left\vert 0\right\rangle +\alpha
_{11}\left\vert 1\right\rangle \left\vert 1\right\rangle )+\left\vert \Phi
_{1}\right\rangle \otimes(\beta_{00}\left\vert 0\right\rangle \left\vert
0\right\rangle +\beta_{10}\left\vert 1\right\rangle \left\vert 0\right\rangle
+\beta_{01}\left\vert 0\right\rangle \left\vert 1\right\rangle )+\left\vert
\Phi_{2}\right\rangle \otimes(\eta_{00}\left\vert 0\right\rangle \left\vert
0\right\rangle +\eta_{10}\left\vert 1\right\rangle \left\vert 0\right\rangle
+\eta_{01}\left\vert 0\right\rangle \left\vert 1\right\rangle )+(\zeta
_{00}\left\vert \Phi_{3}\right\rangle +\chi_{00}\left\vert \Phi_{4}%
\right\rangle +\xi_{00}\left\vert \Phi_{5}\right\rangle )\otimes\left\vert
0\right\rangle \left\vert 0\right\rangle +\prod_{j=1}^{N}\left\vert
g_{j}\right\rangle \otimes(\delta_{20}\left\vert 2\right\rangle \left\vert
0\right\rangle +\delta_{02}\left\vert 0\right\rangle \left\vert 2\right\rangle
)$, here $\left\vert \Phi_{1}\right\rangle =\frac{1}{\sqrt{N}}\sum
\limits_{n=1}^{N}\left\vert e_{n1}\right\rangle \otimes\prod_{j=1,j\neq n}%
^{N}\left\vert g_{j}\right\rangle ,$ $\left\vert \Phi_{2}\right\rangle
=\frac{1}{\sqrt{N}}\sum\limits_{n=1}^{N}\left\vert e_{n2}\right\rangle
\otimes\prod_{j=1,j\neq n}^{N}\left\vert g_{j}\right\rangle $, $\left\vert
\Phi_{3}\right\rangle =\frac{1}{\sqrt{N(N-1)}}\sum\limits_{n=1}^{N}\left\vert
e_{n1}\right\rangle \otimes\sum\limits_{m=1,m\neq n}^{N}\left\vert
e_{m2}\right\rangle \otimes\prod_{j=1,j\neq n,j\neq m}^{N}\left\vert
g_{j}\right\rangle $, $\left\vert \Phi_{4}\right\rangle =\frac{1}%
{\sqrt{N(N-1)}}\sum\limits_{n=1}^{N}\left\vert e_{n1}\right\rangle \otimes
\sum\limits_{m=1,m\neq n}^{N}\left\vert e_{m1}\right\rangle \otimes
\prod_{j=1,j\neq n,j\neq m}^{N}\left\vert g_{j}\right\rangle ,$ and
$\left\vert \Phi_{5}\right\rangle =\frac{1}{\sqrt{N(N-1)}}\sum\limits_{n=1}%
^{N}\left\vert e_{n2}\right\rangle \otimes\sum\limits_{m=1,m\neq n}%
^{N}\left\vert e_{m2}\right\rangle \otimes\prod_{j=1,j\neq n,j\neq m}%
^{N}\left\vert g_{j}\right\rangle $. So the photon loss due to quantum jump
can be calculated by $P_{loss}=1-\left\langle \Psi(t_{I})|\Psi(t_{I}%
)\right\rangle $. \begin{figure*}[ptb]
\includegraphics[width=6.3in]{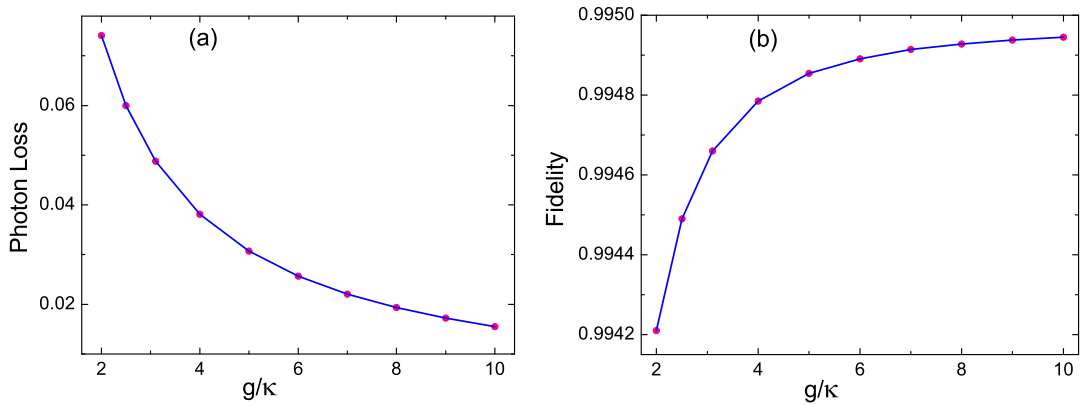}\newline\caption{(Color online) Photon
loss and fidelity of the phase gate vs $g/\kappa$ in (a) and (b),
respectively. Other common parameters: $\kappa=\gamma_{s}$, $g_{ja}=g_{jb}=g$,
$N=4\times10^{4}$, $\Omega=20\sqrt{N}g$.}%
\label{b}%
\end{figure*}Fig. 2 (a) plots the numerical calculation of the photon loss as a
function of $g/\kappa$. The fidelity of the swap gate, which depends on the
initially state of the system, is an efficient measure of the distance between
the quantum logic gates. We define the fidelity (F) of the swap gate as
$F=\left\vert \left\langle \Psi_{ideal}|\Psi_{out}\right\rangle \right\vert
^{2}$, where $\left\vert \Psi_{ideal}\right\rangle =\prod_{j=1}^{N}\left\vert
g_{j}\right\rangle \otimes(\left\vert 0\right\rangle \left\vert 0\right\rangle
+i\left\vert 1\right\rangle \left\vert 0\right\rangle +i\left\vert
0\right\rangle \left\vert 1\right\rangle -\left\vert 1\right\rangle \left\vert
1\right\rangle )/2$ refers to the state of the system in the ideal case after
the gate operation for the initial state $\left\vert \Psi(0)\right\rangle $,
and $\left\vert \Psi_{out}\right\rangle =\left\vert \Psi(t_{I})\right\rangle
/\sqrt{\left\langle \Psi(t_{I})|\Psi(t_{I})\right\rangle }$ since we have
assumed that no quantum jump actually occurs during our implementation of the
scheme \cite{Plenio}. As shown in Fig. 2 (b), the swap gate has high fidelity
even if the strongly coupled condition is not satisfied.

\section{Discussion and conclusion}

In an optical cavity, the coupling strengths $g_{ja},$ $g_{jb}$, and
$\Omega_{j}$ depend on the position \textbf{r} of atom j. However, in presence
of a large number of atoms, due to the many-atom interference effects
resulting in the atoms collectively enhanced coupling to certain optical mode,
it is possible to estimate that even for a high effective coupled parameter
$\xi$ and does not require a strong coupling regime or precise atomic
localization inside a wavelength \cite{Guzman}, which is important for current
experimental technique. In our scheme the atoms are prepared in their ground
states $\prod_{j=1}^{N}\left\vert g_{j}\right\rangle $ and no atomic
transition is required. Thus our scheme is robust against atomic spontaneous emission.

Next we briefly address the experimental feasibility of the proposed scheme.
We consider a low density vapor of $^{85}Rb$ in an optical cavity. The
required atomic level configuration can be chosen from the hyperfine states of
$^{85}Rb$. For instance, the lower level $\left\vert g\right\rangle $ is the
$F=2$ hyperfine state of the $5S_{1/2}$ electronic ground state, while
$\left\vert e_{1}\right\rangle $ and $\left\vert e_{2}\right\rangle $ are the
$F=3$ hyperfine states of the $5P_{3/2}$ electronic excited states,
respectively. Using the cavity parameters of Ref \cite{Hennri}, we have the
waist of the cavity modes $\omega\sim35\mu m$, the waist of the homogeneous
classical laser beam $d\sim50\mu m$, and $(g,\kappa,\gamma_{s})/2\pi
\sim(16,1.4,3)MHz$. With the choices of $N=4\times10^{4}$ and $\Omega
=20\sqrt{N}g$, we obtain an approximate atom number density of $0.8\times
10^{12}/cm^{3}$ (small enough to prevent coherence losses due to collisions),
and the effective coupling constant $\xi\sim10g$. Thus the swap gate operation
time is $t\sim1.6ns$, which is much smaller than the photon lifetime
$t_{c}\sim1/\kappa\sim0.2\mu s$. Hence, our scheme fits well the status of
current experimental technology.

In summary, we have proposed a scheme to realize frequency-up conversion and a
swap gate for intracavity fields with atomic ensemble. In our scheme, the
atoms collectively enhance coupling strength to the intracavity fields due to
the many-atom interference effects. In presence of a large number of atoms,
our scheme may relax the requirements of the strong-coupling condition and
Lamb-Dicke limit, which is important for current experimental technique. We
also discuss the fidelity of the swap\ gate and the probability of the photon
loss under the influence of the atomic spontaneous emission and the decay of
the cavity modes. It is shown that the swap\ gate has a high fidelity and
small error rate.

\textbf{Acknowledgments:} This work was funded by National Natural Science
Foundation of China (Grant No. 10574022), the Natural Science Foundation of
Fujian Province of China (Grant No. 2007J0002 and No. 2006J0230), the
Foundation for Universities in Fujian Province (Grant No. 2007F5041), and
\textquotedblleft Hundreds of Talents \textquotedblright\ program of the
Chinese Academy of Sciences.


\begin{thebibliography}{99}                                                                                               %


\bibitem {Cirac}J. I. Cirac et al., Phys. Scr., T 76, 223 (1998).

\bibitem {Brune}M. Brune et al., Phys. Rev. Lett. 77, 4887 (1996).

\bibitem {Brattke}S. Brattke, B. T. H. Varcoe, and H. Walther, Phys. Rev.
Lett. 86, 3534 (2001); P. Bertet et al., ibid. 88, 143601 (2002).

\bibitem {Rauschen}A. Rauschenbeutel et al., Phys. Rev. A 64, 050301 (R) (2001).

\bibitem {Zheng}S. B. Zheng, Phys. Rev. A 74, 043803 (2006).

\bibitem {Solano}E. Solano et al., Phys. Rev. A 64, 024304 (2001); M. Suhail
Zubairy et al., Phys. Rev. A 68, 033820 (2003).

\bibitem {Garcia}R. Garc\'{\i}a-Maraver et al., Phys. Rev. A 70, 062324 (2004).

\bibitem {Shu}J. Shu et al., Phys. Rev. A 75, 044302 (2007).

\bibitem {Serra}R. M. Serra et al., Phys. Rev. A 71, 045802 (2005).

\bibitem {Zou}X. B. Zou et al., Phys. Rev. A 73, 025802 (2006).

\bibitem {Sondb}A. S\o ndberg S\o rensen and K. M\o lmer, Phys. Rev. A 66,
022314 (2002).

\bibitem {Lukin}M. D. Lukin et al., Phys. Rev. Lett. 84, 4232 (2000).

\bibitem {Dantan}A. Dantan et al., Phys. Rev. Lett. 94, 050502 (2005).

\bibitem {Duan}L. M. Duan et al., Nature (London) 414, 413 (2001).

\bibitem {Josse}V. Josse, A. Dantan, A. Bramati, M. Pinard, and E. Giacobino,
Phys. Rev. Lett. 92, 123601 (2004).

\bibitem {Hennrich}M. Hennrich, A. Kuhn, and G. Rempe, Phys. Rev. Lett. 94,
053604 (2005).

\bibitem {Guzman}R. Guzm\'{a}n, J. C. Retamal, E. Solano, and N. Zagury, Phys.
Rev. Lett. 96, 010502 (2006); A. S. Parkins, E. Solano, and J. I. Cirac, ibid.
96, 053602 (2006).

\bibitem {Xiao}Y. F. Xiao et al., Phys. Rev. A 74, 044303 (2006).

\bibitem {Ottaviani}C. Ottaviani et al., Phys. Rev. A 73, 010301 (R) (2006).

\bibitem {Jeremy}J. Metz et al., Phys. Rev. Lett. 97, 040503 (2006); L. M.
Duan, Phys. Rev. Lett. 88, 170402 (2002).

\bibitem {Zhen}Shi-Biao Zheng, Phys. Rev. A 65, 051804 (R) (2002).

\bibitem {Zubair}M. S. Zubairy, M. Kim, and M. O. Scully, Phys. Rev. A 68,
033820 (2003).

\bibitem {James}D. F. V. James, Fortschr. Phys. 48, 823 (2000).

\bibitem {1}For a review see F. Zernike and I. E. Midwinter, Applied Nonlinear
Optics (Wiley, New York, 1973); R. G. Byer, in Nonlinear Optics, edited by P.
G. Harper and B. S. Wheerett (Academic, New York, 1977).

\bibitem {Barenc}A. Barenco, D. Deutsch, A. Ekert, and R. Jozsa, Phys. Rev.
Lett. 74, 4083 (1995).

\bibitem {Vatan}F. Vatan and C. Williams, Phys. Rev. A 69, 032315 (2004).

\bibitem {Sangouar}N. Sangouard et al., Phys. Rev. A 72, 062309 (2005).

\bibitem {Yavuz}D. D. Yavuz, Phys. Rev. A 71, 053816 (2005).

\bibitem {Plenio}M. B. Plenio and P. L. Knight, Rev. Mod. Phys. 70, 101 (1998).

\bibitem {11}For simplification of the numerical simulations, we assume all
the atoms are in the same situation. In our scheme, this hypotheses is not
prerequisite due to the collective effect of the atomic ensemble (see [17]).

\bibitem {Hennri}M. Hennrich et al., Phys. Rev. Lett. 85, 4872 (2000); P.
Maunz et al., ibid. 94, 033002 (2005).
\end{thebibliography}
\end{document}